\documentclass[structabstract]{aa}
\usepackage{graphicx}
\usepackage{amsmath}
\usepackage{txfonts}
\usepackage{natbib}

\newcommand\lta{\mathrel{\hbox{\raise 0.6 ex \hbox{$<$}\kern
                   -1.8 ex\lower .5 ex\hbox{$\sim$}}}}
\newcommand\gta{\mathrel{\hbox{\raise 0.6 ex \hbox{$>$}\kern
                   -1.7 ex\lower .5 ex\hbox{$\sim$}}}}
 
\newcommand{\scrbox}[1]{\ensuremath{{\mbox{\scriptsize #1}}}}
\newcommand{\teff}{{\ensuremath{T_{\scrbox{eff}}}}}
\newcommand{\Msol}{\ensuremath{\,\mbox{\it M}_{\odot}}}
\newcommand{\Mloss}{\ensuremath{\,\mbox{\it M}_{\odot}{\rm yr}^{-1}}}

\newcommand{\Dturb}{\ensuremath{D_{\scrbox{T}}}}

\newcommand{\MS}{main--sequence}
\newcommand{\gr}{\ensuremath{g_{\scrbox{rad}}}}

\newcommand{\DM}{\ensuremath{ \log \Delta M/M_{*}}}

\renewcommand{\H}{\mbox{H}}
\newcommand{\He}{\mbox{He}}

\newcommand{\Fe}{\mbox{Fe}}

\newcommand{\Mg}{\mbox{Mg}}

\newcommand{\Ca}{\mbox{Ca}}
\newcommand{\Ni}{\mbox{Ni}}

\newcommand{\Ti}{\mbox{Ti}}

\newcommand{\Li}{\mbox{Li}}

\newcommand{\T}{_{\rm{T}}}

\begin{document}

\title{Population II stars and the Spite plateau}
\subtitle{Stellar evolution models with mass loss}

\author{M. Vick\inst{1,}\inst{3}, G. Michaud\inst{2,}\inst{3}, J. Richer\inst{3}, 
 \and O. Richard\inst{1}}

\institute{LUPM, Universit\'e Montpellier II, CNRS
                 CC072, Place E. Bataillon,
                 34095\,Montpellier Cedex,
                 France
		\and
            LUTH, Observatoire de Paris,
	    CNRS, Universit\'e Paris Diderot,
	    5 Place Jules Janssen, 92190\,Meudon, France           
\and
	   D\'epartement de physique, Universit\'e de Montr\'eal, 
           Montr\'eal, Qu\'ebec, H3C 3J7, Canada\\
           \email{mathieu.vick@umontreal.ca, olivier.richard@univ-montp2.fr,
	   michaudg@astro.umontreal.ca,
           jacques.richer@umontreal.ca}
}

\date{\today}

\abstract{}
{We aim to determine the constraints that observed chemical abundances put on the potential role of mass loss in metal poor dwarfs.}
{Self-consistent stellar evolutionary models that include all the effects of atomic 
diffusion and radiative accelerations for 28 chemical species were 
computed  for stellar masses between 0.6 and 0.8\Msol{}.  Models with an initial metallicity of $Z_0=0.00017$ and mass loss rates from $10^{-15}\Mloss$ to $10^{-12}\Mloss$  were calculated.
They were then compared to previous
models with mass loss, as well as to models with turbulent mixing.}
{For models with an initial metallicity of $[\Fe/\H]_0=-2.31$, mass loss rates of about 
$10^{-12}\Mloss{}$ lead to surface abundance profiles that are very similar to those obtained in models with 
turbulence. Both models  have about the same level of agreement with observations of galactic-halo lithium abundances, as well as
lithium and other elemental abundances from metal poor globular clusters such as NGC\,6397.
In this cluster, models with mass loss agree slightly better with subgiant observations of Li abundance than those with turbulence.  Lower red giant branch stars instead favor the models with turbulence.  Larger differences between models with mass loss and those with turbulence appear in the interior concentrations of metals.   }
{The relatively high mass loss rates required 
to reproduce 
plateau-like lithium abundances appear unlikely when compared to the solar mass-loss rate.  However the presence of a chromosphere on these stars justifies  further investigation of the  mass-loss rates.   }


\maketitle{}

\section{ASTROPHYSICAL CONTEXT}
\label{sec:intro}

Population II stars can be used as tracers for exploring the universe's chemical evolution.  
Still, to correctly relate the chemical properties of the 
early Universe to the composition of the oldest observed stars, one must  
be able to map the evolution of their current abundances back to  their original values. 
Doing so could help elucidate numerous questions pertaining to Big Bang nucleosynthesis, the 
nature of the first supernovae, and the age of the Universe.

The nearly constant Li abundance of most low-metallicity ([Z/H]\,$<$-$1.5$) halo field stars whose
\teff{} is between 6300\,K and 5500\,K -- otherwise known as the Spite plateau 
(\citealt{spite82,spite84}) -- 
has puzzled astronomers for decades. Originally, many inferred that this plateau was 
representative of the Universe's original Li abundance, though it was 
promptly shown that this could not be
the case (\citealt{michaud84}).
Precise observations of the cosmic microwave background\footnote{From the \emph{Wilkinson 
Microwave Anisotropic Background}, WMAP, see \citealt{spergel07}.} have now
led to a primordial lithium abundance\footnote{$A(\Li)=\log[N(\Li)/N(\H)+12]$.}  of
$A(^7\Li)$$=2.65_{-0.06}^{+0.05}$,
 or even higher, $A(^7\Li)=2.72_{-0.05}^{+0.05}$ \citep{cyburt03,steigman07,cyburt08}, while 
the lithium content in metal poor dwarfs, ranging from 
$A(^7\Li)=2.0$ to $2.4$  is a factor of $2-3$ lower \citep{bonifacio07462,asplund06, charbonnel05primas,
melendez04,bonifacio02,ryan99}.

Changing the standard 
Big Bang nucleosynthesis model is a possible explanation, though it is 
hypothetical (\citealt{ichikawa04,coc04,cumberbatch07,RegisCl2012}).
Even though Population III stars may have destroyed
some of the primordial lithium (\citealt{piau06}), this cannot explain the large discrepancy.
This leads many to 
assume that the solution 
stems from within the stars themselves. Possible stellar explanations may 
include destruction via gravity waves 
(\citealt{talon04,charbonnel05science}), and rotational mixing 
(\citealt{vauclair88,charbonnel92,pinsonneault99}), 
as well as settling via atomic diffusion 
(\citealt{deliyannis90,profitt91,salaris00,RichardMiRi2005}). 
A higher Li content of subgiant (SG) stars compared to \MS{} (MS) 
stars in globular clusters (GC) such as NGC 6397 
(\citealt{korn07, lind09NGC6397})
favors a model in which  lithium is not destroyed, but simply sunk 
below the photosphere prior to dredge-up.  This result is, however, sensitive to the \teff{} scale used (\citealt{hernandez09,NordlanderKoRietal2012}).

Additional constraints may stem from claimed observations of $^6\Li$ (\citealt{asplund06}), 
which,
with a thermonuclear destruction cross section 60 times larger than the heavier isotope, 
would severely constrain the stellar destruction of $^7\Li$. These observations 
remain very uncertain
(\citealt{cayrel07,LindAsCoetal2012}). Many recent studies have also
claimed that below [Fe/H]$ \simeq -2.5$, Li
abundances lie significantly below the Spite plateau abundance and with increased scatter 
(\citealt{asplund06}, 
 \citealt{aoki09}, \citealt{SbordoneBoCaetal2010}, \citealt{SbordoneBoCa2012}). Other observations
suggest that there might instead be a lower plateau abundance for extremely metal poor (EMP) 
stars (\citealt{melendez10}). The situation is complex and not currently understood in detail.

Although the surface convection zone (SCZ) of galactic Population II 
stars homogenizes abundances 
down to depths at which atomic 
diffusion is relatively slow, the long lifetimes of these stars, which 
reach the age of the Universe, allows for
the effects of chemical separation to materialize at the surface. 
However, models that allow atomic diffusion to take place unimpeded 
lead to lithium underabundances that
are too large at the hot end of the plateau (\citealt{michaud84}).
By enforcing additional turbulent mixing below the SCZ, models of 
metal poor stars with gravitational 
settling and radiative accelerations are able to reproduce many 
observations of halo and globular cluster stars \citep{richard02I, 
VandenbergRiMietal2002,richard02III,RichardMiRi2005,korn06}. These same models were also 
successful in explaining abundance anomalies in AmFm (\citealt{richer00,richard01}) and 
horizontal branch stars (\citealt{michaud07,michaud08}).

In \citet{vauclair95}, gravitational 
settling is inhibited by  unseparated  
 mass loss at the constant  rate of \hbox{$\sim 3 \times 10^{-13}\Mloss$}; this leads to near constant 
lithium abundances as a function of \teff{}, even
at the hot end of the plateau. However, these calculations, which are discussed in 
Sect.\,\ref{sec:vauclair}, did not include the effects of radiative
accelerations, nor did they follow atomic diffusion in detail for elements heavier than Li; 
therefore, further inspection is warranted.


In \citet{vick10,vick11}, hereafter Papers I and II, 
stellar evolution models with
mass loss were introduced and shown to reproduce observed 
surface abundance anomalies for many AmFm and chemically anomalous HAeBe 
binary stars. In \citet{MichaudRiVi2011} it was then shown that a mass loss rate compatible with the observed Sirius A mass loss rate leads to abundance anomalies similar to those observed on Sirius A. To further validate the model and to help constrain the relative 
importance of mass loss
with respect to turbulent mixing in stars, the present 
mass loss models will be applied to Pop II stars and compared to models 
with turbulence from \citet{richard02I} and \citet{RichardMiRi2005},  
which are able to reproduce
observations of galactic halo stars, as well as of globular cluster stars on the subgiant and giant branches. 


In the following analysis and in our calculations, 
mass loss is considered in 
\emph{non rotating stars}, 
since in these stars it might be the only process competing with atomic diffusion. 
To our knowledge, mass loss  has never 
been observed
in Population II MS stars; therefore, mass loss rates
will be constrained solely via surface abundance anomalies. In 
Sect.\,\ref{sec:calcul}, the evolutionary calculations are presented, 
while the results 
are discussed in Sect.\,\ref{sec:models}. In Sects.\,\ref{sec:vauclair} 
and \,\ref{sec:turbulence}, the models with mass
loss are compared to previous evolutionary models of metal poor dwarfs with mass loss, 
as well as to models with turbulence. In Sect.\,\ref{sec:Subgiant} results are compared to subgiant and giants both for  models with turbulence and with mass loss. Conclusions  follow in
Sect.\,\ref{sec:conclusion}.

\section{Calculations}
\label{sec:calcul}
The stellar evolution models were calculated
as explained in Paper I and references therein. 
Atomic diffusion is computed self-consistently 
with radiative 
accelerations from \citet{richer98} and corrections for redistribution from
\citet{gonzalez95} and \citet{leblanc00}. Chemical transport is computed 
for 28 chemical species
(all those included in the OPAL database, \citealt{iglesias96}. plus Li, Be, and B). The models are 
evolved from the pre--main-sequence (PMS)
with the abundance mix prescribed in \citet{richard02I} for $Z_0=0.00017$ 
(or equally, [Fe/H]$=-2.31$). 
Most models were calculated with an initial helium fraction of 0.235 (corresponding to the $Y_0$ used for calculations for M\,92 in \citealt{VandenbergRiMietal2002}). Since one of our major aims is to compare with similar models calculated with turbulence, this value was chosen to facilitate comparison (since it is the same as in \citealt{richard02I} and \citealt{RichardMiRi2005}). Some models were also calculated with $Y_0=0.2484$, which was derived from {WMAP} observations \citep{cyburt03} and is discussed in \S\,3.3.1 of \citet{RichardMiRi2005}. As shown in Fig. 4 of \citet{VandenBergBoSt96} and Fig. 5 of \citet{RichardMiRi2005}, respectively, this small change in initial helium fraction has little effect on the luminosity--versus--age relation{$^{\ref{fn:Y2484}}$} or on lithium isochrones.  
With 
respect to the solar abundance mix
used in Papers I and II for Population I stars, the relative abundances of alpha
elements are increased as appropriate for Population II stars (\citealt{vandenberg00}).
The introduction of mass loss and its impact on transport
is outlined in Paper I. 

The convection zone boundary is determined by the Schwarzchild criterion with energy flux in the convection zone computed with  \citet{BohmVitense92} mixing length formalism. Abundances are assumed to be completely mixed within convection zones.
No extra mixing is enforced outside of convection zones. 
The effects of atomic diffusion materialize as soon as radiative 
zones appear as the models evolve toward the MS. These effects 
become more important as the radiative zone expands toward the surface, where 
atomic diffusion becomes more efficient, and the convection zone becomes less massive, 
thereby reducing the dilution of anomalies.

The unseparated\footnote{In this context, \emph{unseparated} refers to the abundances 
in the wind being the same as in the photosphere.} mass loss 
rates considered range from $10^{-15}$ 
to $10^{-12}$\Mloss. 
Owing to uncertainties related to the nature of 
winds for Population II stars (see also discussion in Sect.\,4.2 of Paper I), we 
have chosen to limit our 
investigation to unseparated winds
in order to avoid introducing additional adjustable parameters.

In order to fully constrain the effects of mass loss, we have chosen to use the 
same boundary condition (\citealt{krishna66}) and mixing length parameter, $\alpha = 2.096$, as in
Papers I and II (see also Sect.\,2 of Paper I and \citealt{vandenberg08}). Most models 
in \citet{richard02I,richard02III} and \citet{RichardMiRi2005} were computed with the Eddington gray atmosphere surface 
boundary condition (see also \citealt{turcotte98soleil}).
The main consequence is to change the value of $\alpha$. In both cases, the mixing 
length is calibrated using the Sun so that very little difference remains 
in solar type models.
Therefore, though the different boundary conditions might lead to a 
different evolution of SCZ depths and masses 
(see Fig.\,9 of \citealt{richard02I}), 
the remaining slight difference does not play a role in models with turbulence since 
the separation that modulates surface abundances usually occurs deeper. As shown
in Sect.\,\ref{sec:models}, the same is true for the models with mass loss presented in this 
paper since separation also occurs deeper than the bottom of the convection zone. 
\begin{figure*}[!t]
\begin{center}
\includegraphics[scale=0.95]{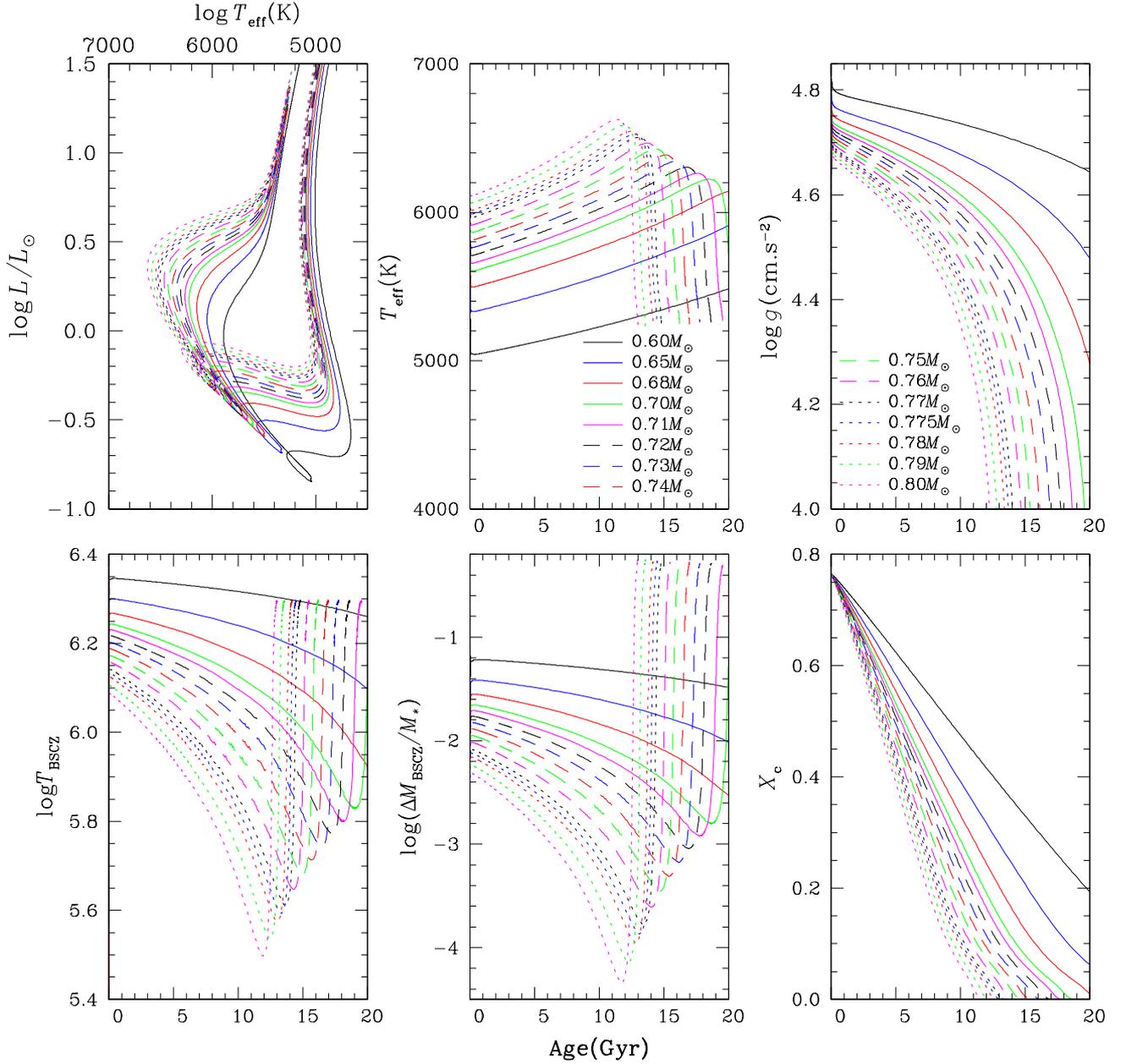}
\caption{The Hertzprung-Russel diagram, as well as the evolution of \teff{}, $\log g$, the
temperature at the base of the SCZ ($T_{{\rm BSCZ}}$), the outer mass at the base of the SCZ ($ \Delta M_{{\rm BSCZ}}$),
and the mass fraction of hydrogen in the core ($X_{\rm c}$) 
for stars of $0.6-0.8$\Msol{} with $Z_0=0.00017$ and 
${\dot M}=10^{-12}\Mloss$. 
}\label{fig:HR}
\end{center}
\end{figure*}
\begin{figure*}[!t]
\begin{center}
\includegraphics[scale=0.95]{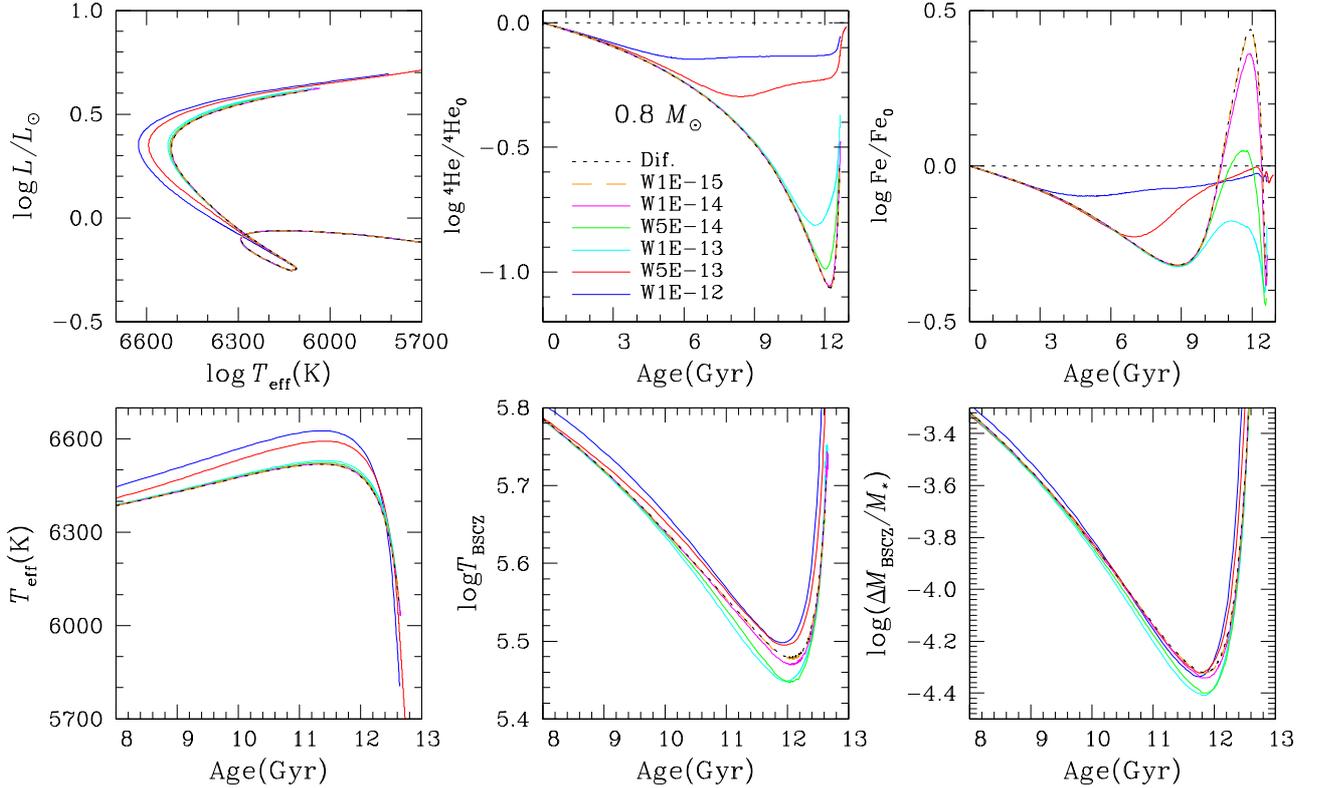}
\caption{The impact of mass loss  on the evolution of the depth (in temperature and mass) of 
the SCZ in 0.8\Msol{} models with atomic diffusion only, and with mass loss rates from $10^{-15}\Mloss$ to $10^{-12}\Mloss$.  \emph{upper row:} HR diagram and the \He{} and \Fe{} surface abundances. \emph{lower row:} Zoom on the \teff{}, and on the mass and temperature at the bottom of the SCZ. 
}\label{fig:histmdot}
\end{center}
\end{figure*}

\section{Evolutionary models}
\label{sec:models}

Evolutionary models of masses ranging from $0.6$ to $0.8$\Msol{}, an interval of masses 
that spans the Spite
plateau, are shown in a Hertzprung-Russel diagram (H-R) in Fig.\,\ref{fig:HR}. All the 
models have
an initial metallicity of $Z_0=0.00017$, and a mass loss rate of $10^{-12}\Mloss$. 
Figure\,\ref{fig:HR} also shows the evolution of \teff{}, gravity, the depth 
(in mass, $\Delta M_{{\rm BSCZ}}$, 
and in temperature, $T_{{\rm BSCZ}}$) of the bottom of the SCZ, and
the central H mass fraction. 
The 0.79 
and 0.77\Msol{} models are respectively at the MS turnoff at 12\,Gyr and 13.5\,Gyr, the assumed lower and upper limits 
of the age of halo dwarfs.  

For all models, the variations in surface convection zone depth have an 
impact on surface--abundance
evolution since the efficiency of atomic diffusion varies 
with depth. Throughout the MS, $\Delta M_{{\rm BSCZ}}$ decreases 
until H is exhausted in the stellar core (i.e. until turnoff). 
Although this and other parameters depend on the mass loss rate, the effect is slight 
(see Fig.\,\ref{fig:histmdot}). Just before the MS turnoff, there is a 
$\sim 125\,$K difference in \teff{} between the 0.8\Msol{} model 
with \textit{diffusion only}
and the \hbox{0.80W1E-12 model}\footnote{0.80W1E-12 refers to a 0.80\Msol{} model with 
a mass loss rate of 
$1 \times 10^{-12}\Mloss$.}, while the difference falls to about 15\,K between the 
\textit{diffusion only} and 0.80W1E-13 models. The temperature and mass at the bottom of the SCZ are related to both the \He{} and \Fe{} concentrations in the SCZ.  For instance, the temperature  is lowest when both \He {} and \Fe{} have a small abundance at turn--off in the 0.80W1E-13 model.  The temperature is highest in the models with largest \He {} and \Fe{} abundances (and highest mass loss rate) and intermediate in the diffusion only model where the low \He{} abundance is partly canceled by the high \Fe{} abundance.  The iron abundance at turn--off is largest (+0.47\,dex) in the model without mass loss and  decreases progressively to -0.2 dex as the mass loss rate increases to $10^{-13}\Mloss$ and then is very nearly normal in the larger mass loss cases.
The effect on stellar 
age at the turnoff, or equivalently, on the stellar 
lifetime, is less than the line thickness in the plots.
While $\Delta M_{{\rm BSCZ}}/M_*$ varies 
by a factor of 100 ($\DM$ goes from $\sim -2.3$ to 
$-4.3$) on the MS for a 0.80\Msol{} model (Fig.\,\ref{fig:HR}), the maximum effect 
of adding mass loss is below 0.08\,dex. 
 
\subsection{Radiative accelerations, mass loss rate, and internal abundance variations}
\label{sec:abint}
\begin{figure*}[!t]
\begin{center}
\includegraphics[scale=0.95]{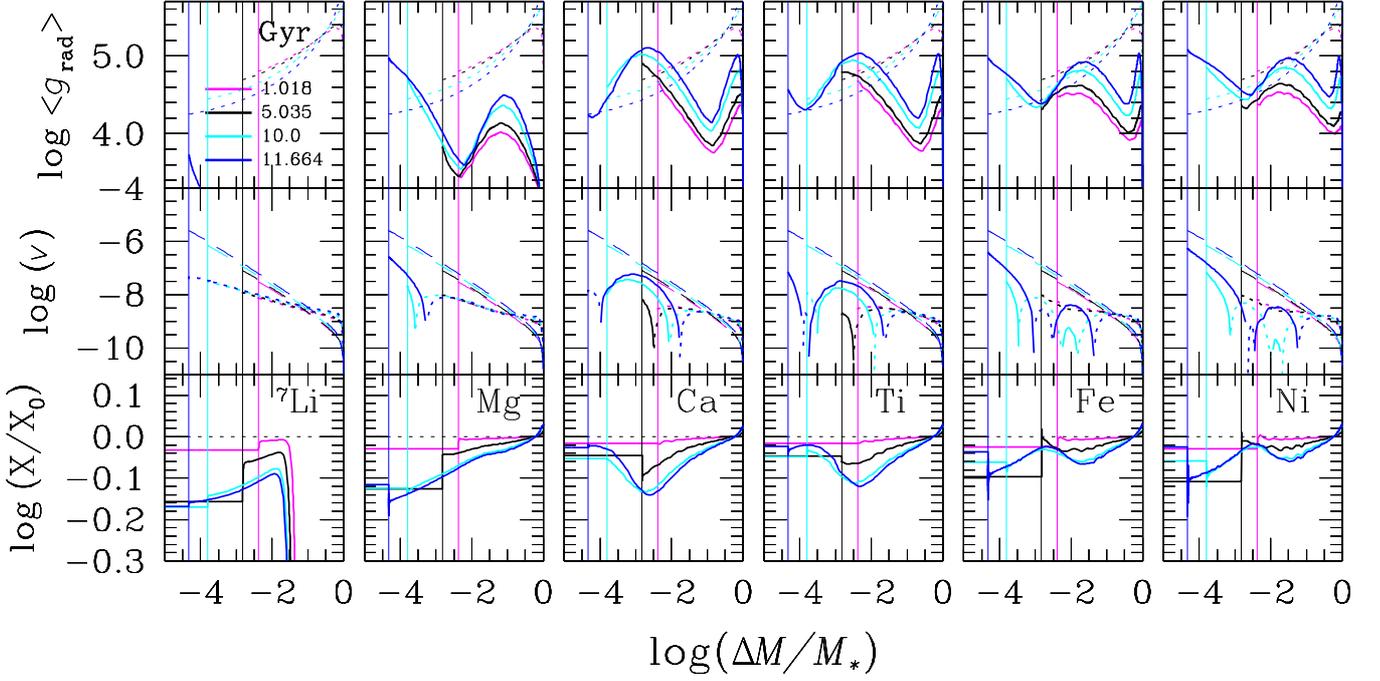}
\caption{\emph{upper row:} Radiative accelerations (solid lines) and gravity (dotted lines); 
\emph{center row:}Wind velocities (dashed lines) and diffusion velocities toward the exterior   (solid lines) and toward the interior (dotted lines);
{\emph{lower row:} Interior concentrations as a fraction of original ones.  
These quantities are shown at four 
different ages for a 0.8\Msol{} model with 
${\dot M}=10^{-12}\Mloss$. The vertical lines indicate 
the position of the bottom of the SCZ. The ages are identified in the figure.}}
\label{fig:temporel}
\end{center}
\end{figure*}
\begin{figure*}[!t]
\begin{center}
\includegraphics[scale=0.95]{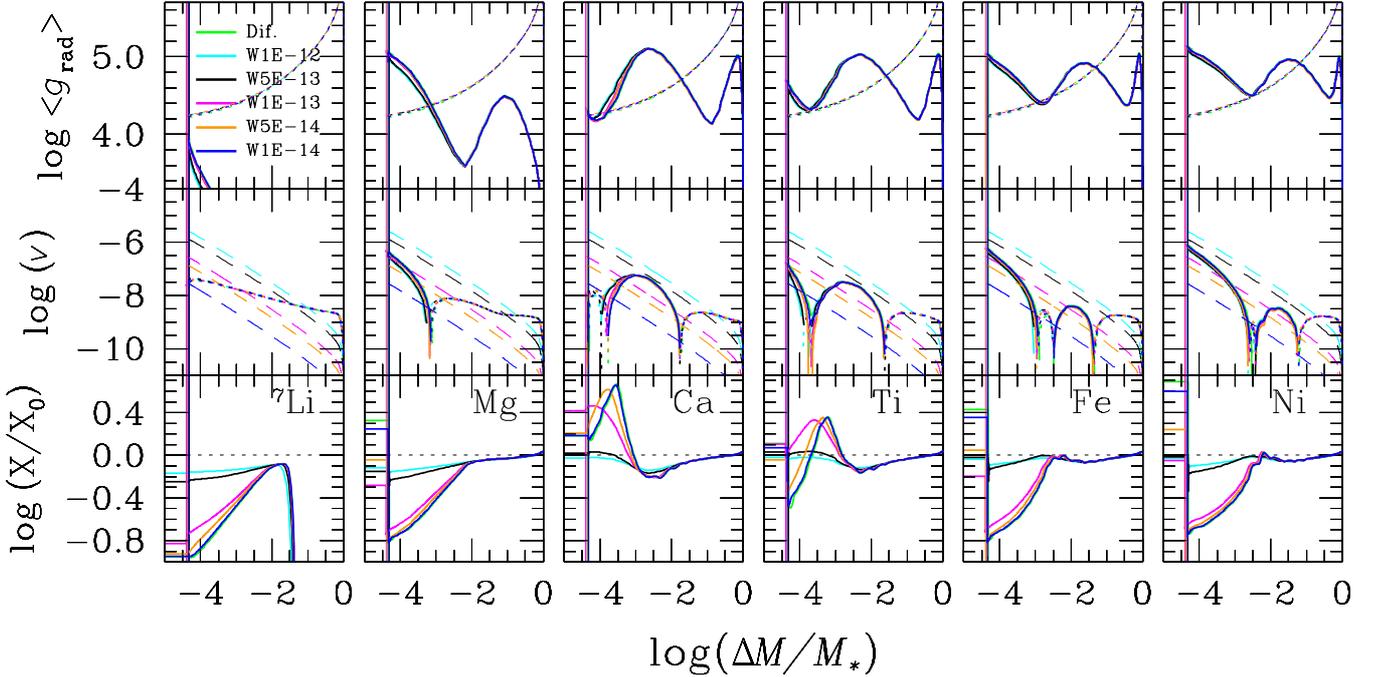}
\caption{The impact of  mass loss rate on internal abundance 
profiles for 0.80\Msol{} models. A model without mass loss and models with five mass loss rates from $10^{-14}\Mloss$ to $10^{-12}\Mloss$ are shown and are identified in the figure. The internal concentration profiles are shown just before 
turnoff near 11.5\,Gyr. Atomic diffusion
affects mainly the outer $10^{-2}$ of the stellar mass. Conventions are similar to those of Fig.\,\ref{fig:temporel}.
}\label{fig:mdot}
\end{center}
\end{figure*}
On very short time scales, convection homogenizes abundances from 
the bottom of the SCZ to the 
stellar surface; therefore, 
the competition between radiative acceleration (\gr) and gravity is only important below 
the surface convection zone.  In Fig.\,\ref{fig:temporel},
the internal variations of \gr{}, velocity, and mass fraction are shown for  \Li, \Mg, \Ca, \Ti, \Fe{}, and \Ni{} in 
the 0.80W1E-12 model\footnote{This mass
was chosen  to facilitate comparison with the models and figures
from \citet{richard02I}.}. Having \gr{}, velocity, and mass fraction aligned for a given element facilatates the discussion of the link between those quantities. Figures for \gr{} and mass fraction of all elements are shown in App.\,\ref{sec:Appendix}; they allow to show the systematic variations of those quantities. The changes in \gr{} over time 
are due principally to structure changes, since the abundances 
in Population II stars with $[\Fe/\H]_0=-2.31$ are never large enough to cause
saturation\footnote{The mass loss rate also has little to no 
effect on \gr{} for this same reason, and because its
effects on the structure are very small (see Fig.\,\ref{fig:histmdot}).}. 

The 
internal variations of an element's \gr{} 
can be related to its
ionization state as may be seen in the online App.\,\ref{sec:Appendix}; H-like electronic configurations 
lead to local maxima, while He-like configurations 
lead to local minima. As we 
move toward elements of higher atomic number, the H-like state is encountered at greater depths. 
Carbon for instance, is in an H-like configuration
at $\DM\sim -3$, while for Si, this occurs near $\DM\sim -1$. Some elements, including Fe and Ni,
never quite reach the H-like state\footnote{To be more precise, in the stellar core, these
elements arrive at an equilibrium between multiple ionization states, including the hydrogenic one.}. Unlike the H-like maximum, 
the second local \gr{} maximum, which is clearly distinguishable for all 
elements heavier than S, is 
usually greater than or equal to gravity. This \gr{} increase corresponds 
to configurations between 
F-like and Li-like. 

For many elements, including 
He, Li, and O, the radiative acceleration is always smaller 
than gravity. For a mass loss rate of  $10^{-12}\Mloss{}$, these elements are underabundant (bottom row of Fig.\,\ref{fig:temporel}) throughout the star except close to the center. 
Other elements such as Be, Mg, and Fe have
local \gr{} maxima just below the SCZ,  
which, at 11\,664\,Myr, causes their surface abundances to be larger than just below the convection zone.    
 All abundance
variations caused by atomic diffusion are smaller than 0.2\,dex 
(in contrast to the ones caused by thermonuclear 
reactions which are encountered near and around 
the stellar core for $Z < 8$). For this mass loss rate, all elements are 
underabundant below the 
SCZ, and down to about $\DM \sim -0.5$. What varies most rapidly is the location of the bottom of the SCZ.  
The time variation of  interior concentrations is continuous and relatively small.  Larger variations are linked to local extrema of \gr{}. For instance $X(\Fe)$ has a local maximum at $\DM \sim -3$ where $\gr(\Fe)$ has a minimum and the diffusion velocity is inwards. These effects will be  discussed below in relation to the effect of varying the mass loss rate.
Below $\DM \sim -0.5$, elements heavier than O 
become slightly overabundant due to gravitational settling. Therefore, for these 
elements, there are
underabundances of up to a factor of 1.6 over the outer half of the stellar mass and 
overabundances of a few percent 
over the inner half. 

For 
models with lower  mass loss rates, internal 
abundance variations become much greater (Fig.\,\ref{fig:mdot}). 
In the 0.80\Msol{}  model with a $10^{-13}\Mloss{}$ mass loss rate, \Ca{} and \Ti{} are locally overabundant by a factor of up to $\sim$ 3, while $^4$He, $^7$Li, \Mg{}, and \Fe{}
 become $-$0.8\,dex 
underabundant. These variations can be 
linked directly to variations in \gr. For example, the Ni underabundance that 
begins at $\DM\sim -2.5$, and reaches up to the BSCZ, is caused by the 
increase in its \gr{} over the same
depth interval. In this region, Ni is pushed 
into the SCZ, thus causing the Ni abundance to be relatively larger at the surface and to
decrease below the SCZ. 
In contrast, Ca is  
overabundant near $\DM= -3.5$ and above,  since \gr(Ca) has a local 
maximum just below this depth.

In fact, when mass loss is added, one may distinguish  two 
types of internal solutions: 
those dominated by atomic diffusion and those dominated by 
mass loss (see also Sect.\,5.1.1 of 
Paper I). For any given model with mass loss, 
an element's internal abundance solution 
is determined by the mass loss rate 
wherever the associated velocity, $v_{{\rm wind}}$, is greater in 
amplitude than the settling
velocity $v_{{\rm sett}}$. For example, in Fig.\,\ref{fig:mdot}, 
one first notes that the depth at which $|v_{{\rm wind}}|\simeq|v_{{\rm sett}}|$, which we 
 call the point of
separation, depends on the
element and on the stellar age (through structure changes). For the 
0.80W1E-12 model, mass loss
dominates the internal solution of $^7\Li$ from the surface 
down to $\DM\sim-1.2$, while this
occurs nearer $\DM\sim-0.9$ for Ni. As discussed in 
Sect.\,5.1.1 of Paper I, the 
consequence is that local abundances above the point of separation are not 
determined uniquely by the local \gr{}, but 
instead adjust in order to conserve the 
flux arriving from just below the point of separation.
If the wind--dominated solution extends down to 
$\Delta M \equiv {\dot M} \times t$, the surface abundance 
reflects $\gr/g$ at $\Delta M$.
If we come back to the previous example, at 10\,Gyr---or $10^{10}$yr--- and for a mass loss
rate of $10^{-13}\Mloss$, the surface  
reflects $\gr/g$ at $\sim 10^{-3}M_*$.

In general, a higher mass loss rate has the effect of reducing internal anomalies. 
For example, a mass loss rate of 
 $10^{-12}\Mloss{}$ reduces the \Li{} and Mg anomalies below the convection zone  by $\sim$0.6\,dex (Fig.\,\ref{fig:mdot}).
However, the abundance anomalies of, for instance, \Fe{} do not always have a monotonous variation with mass loss rate.  As one reduces the mass loss from  $10^{-12}\Mloss$, the \Fe{} abundance in and below the SCZ first decreases and is 0.6\,dex smaller below the SCZ in the $10^{-13}\Mloss$ model.  In the SCZ it is 0.2\,dex smaller.  However, in the model with a mass loss rate of $10^{-13}\Mloss$, the surface \Fe{} abundance is 0.35\,dex larger than the original abundance, and without mass loss, it is 0.4\,dex larger than the original one\footnote{We have verified that the $10^{-15}\Mloss$ model is the same as the no--mass--loss model.  See Fig.\,\ref{fig:histmdot}.}.  This surprising behavior is related to the minimun of \gr(\Fe) at $\DM \sim -3$.  For a mass loss rate of $10^{-13}\Mloss$, the wind carries away the surface \Fe{} pushed by \gr(\Fe) above $\DM \sim -3$ since in one Gyr, it carries $\DM \sim -4$ and the SCZ is slightly less than  $\DM \sim -4$.  The \Fe{} flux that crosses the region where the \Fe{} diffusion velocity is negative, at $\DM \sim -3$,  is reduced, leading to the observed underabundances.  The $10^{-14}\Mloss$ wind carries a lower \Fe{} mass from the outer region and reduces the surface \Fe{} by only 0.05\,dex compared to the no--mass--loss model. 

On the other hand,  Ca in the
\hbox{0.80W1E-13 model} in Fig.\,\ref{fig:mdot} has a larger surface overabundance than in the no--mass--loss model.  This is because the \Ca{} abundance must adjust to carry the flux strongly pushed upwards at  $\DM \sim -3$ where $\gr(\Ca)$ is much larger than gravity.  In the $10^{-14}\Mloss$ model, the wind velocity is too weak to carry the flux when the diffusion velocity is toward the interior, and there is a sharp peak of $X(\Ca)$ at $\DM = -3.5$. The profile of the concentration is clearly linked to the profile of the downward oriented diffusion velocity, from \Mg{} to \Fe{}.
\begin{figure*}[!t]
\begin{center}
\includegraphics[scale=1.0]{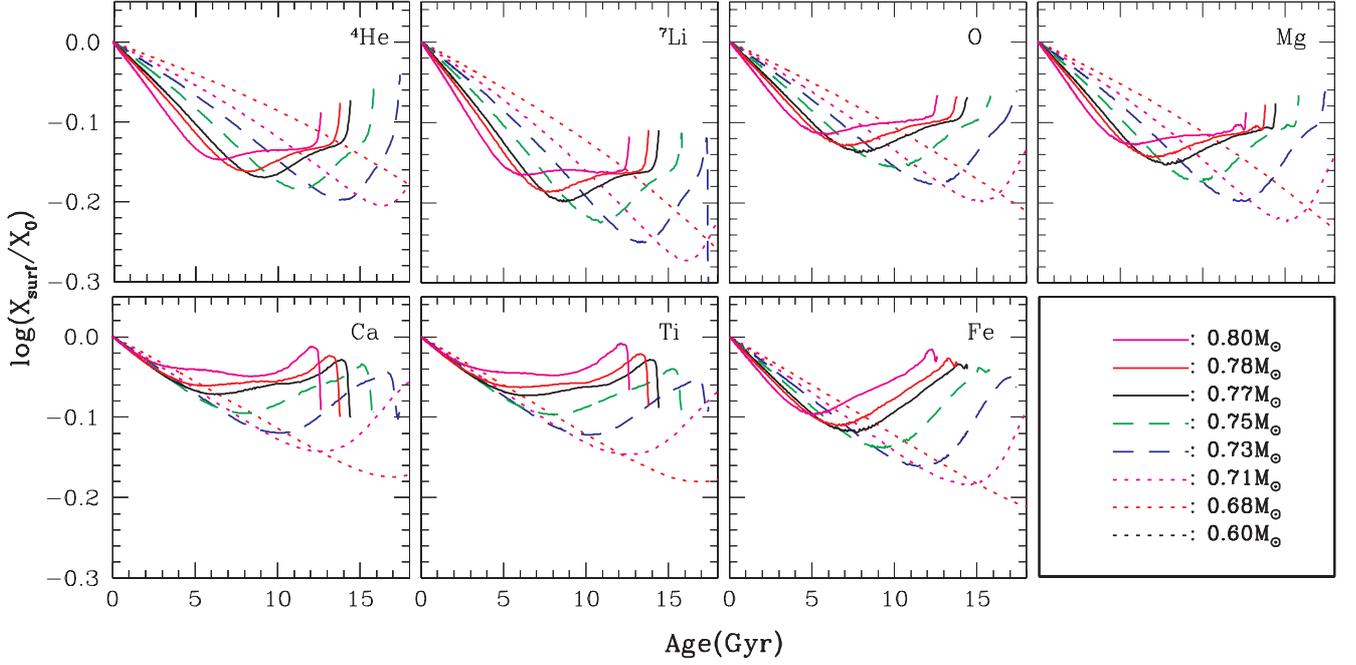}
\caption{Evolution of surface abundance anomalies for 
selected stellar masses and elements with a mass loss rate of $10^{-12}\Mloss$.
}\label{fig:absurfM}
\end{center}
\end{figure*}

\subsection{Surface abundance variations}
\label{sec:absurf}

For a given metallicity, surface abundances depend 
on stellar mass and age as 
well as on  mass loss rate. The age and stellar mass 
dependance are  demonstrated in Fig.\,\ref{fig:absurfM}, which shows 
the evolution of
surface abundances for selected models with a mass loss rate of $10^{-12}\Mloss$. 
First, for this mass loss rate, all elements are underabundant throughout 
MS evolution, and 
until dredge up advects layers which are affected by thermonuclear reactions. The 
underabundance amplitudes increase as stellar mass decreases, since settling becomes 
more important for
models with longer lifetimes. Over a large fraction of the stellar lifetime, the 
overall \textit{shape} of the surface abundance solution is quite similar for all 
masses since the overall shape of \gr{} does not vary significantly over this mass
interval.  See Fig.\,\ref{fig:histmdot} for the effect of varying the mass loss rate on the \He{} and \Fe{} abundances.
   


\begin{figure}
\begin{center}
\includegraphics[scale=0.60]{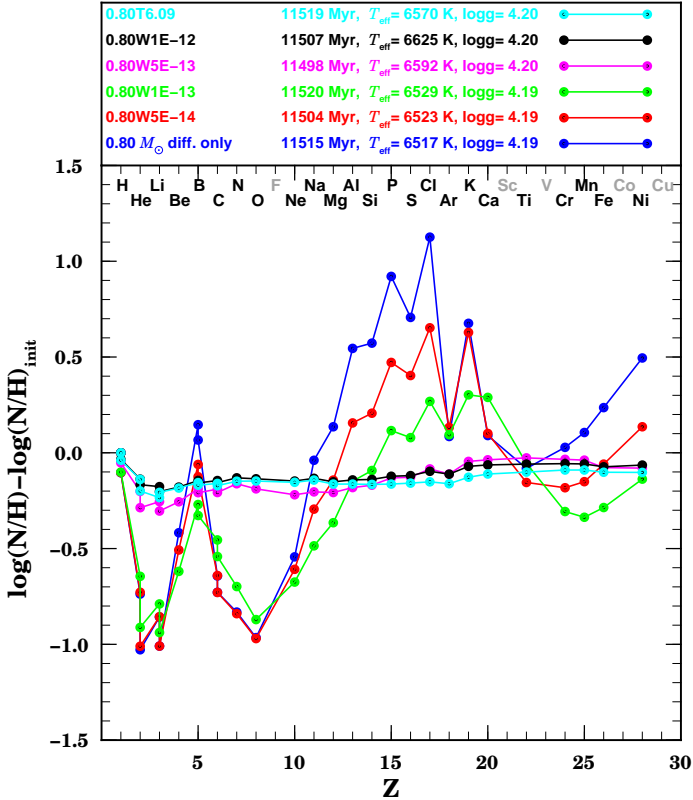}
\caption{Surface abundance anomalies for 0.8\Msol{} models
with and without mass loss as
well as with the T6.09 turbulent parametrization. At 11.5\,Gyr, these 
models are just before turnoff. 
}\label{fig:absurfmdot}
\end{center}
\end{figure}

At a given age, the mass loss rate  affects the surface abundance of the various elements differently, since supported 
elements are expelled through the
surface, while sinking elements encounter advection by $v_{{\rm wind}}$.
For 0.80\Msol{} models near turnoff (Fig.\,\ref{fig:absurfmdot}), a mass loss rate of 
$10^{-13}\Mloss$ reduces He, Li, and O underabundances obtained in the 
\textit{diffusion only} model by a little more than 0.1\,dex,
while also reducing overabundances of Cl and P by about 0.8\,dex. It is then clear that for this mass loss rate, the wind's effect is greater for overabundant (i.e. supported) elements, since the settling velocity still dominates for unsupported elements. For yet other
elements the abundance variations are more complex: the overabundances of 
Fe and Ni become underabundances,
while Ca and Ti become more overabundant for a short 
period of time. For Ca and Ti, this is because $v_{{\rm wind}}$ has an effect just 
below the SCZ, where 
$\gr{}\simeq g$ (Fig.\,\ref{fig:mdot}); therefore, instead of accumulating 
below the SCZ, they are pushed into it. Similarly, 
mass loss advects the layers just below the SCZ, where Fe and Ni 
abundances are smallest, into the SCZ (see Fig.\,\ref{fig:mdot} and \ref{fig:allX_Mloss}). For the two higher 
mass loss rates, the surface anomalies are reduced to factors mostly below about 0.2\,dex. 
While the \hbox{0.80W1E-12} has generalized underabundances, the \hbox{0.80W5E-13} model has small
overabundances of elements from K to Mn (see Fig.\,\ref{fig:allX_Mloss}). 
\label{sec:lithium}

\subsubsection{Surface lithium abundances}

\begin{figure*}[!t]
\begin{center}
\includegraphics[scale=0.90]{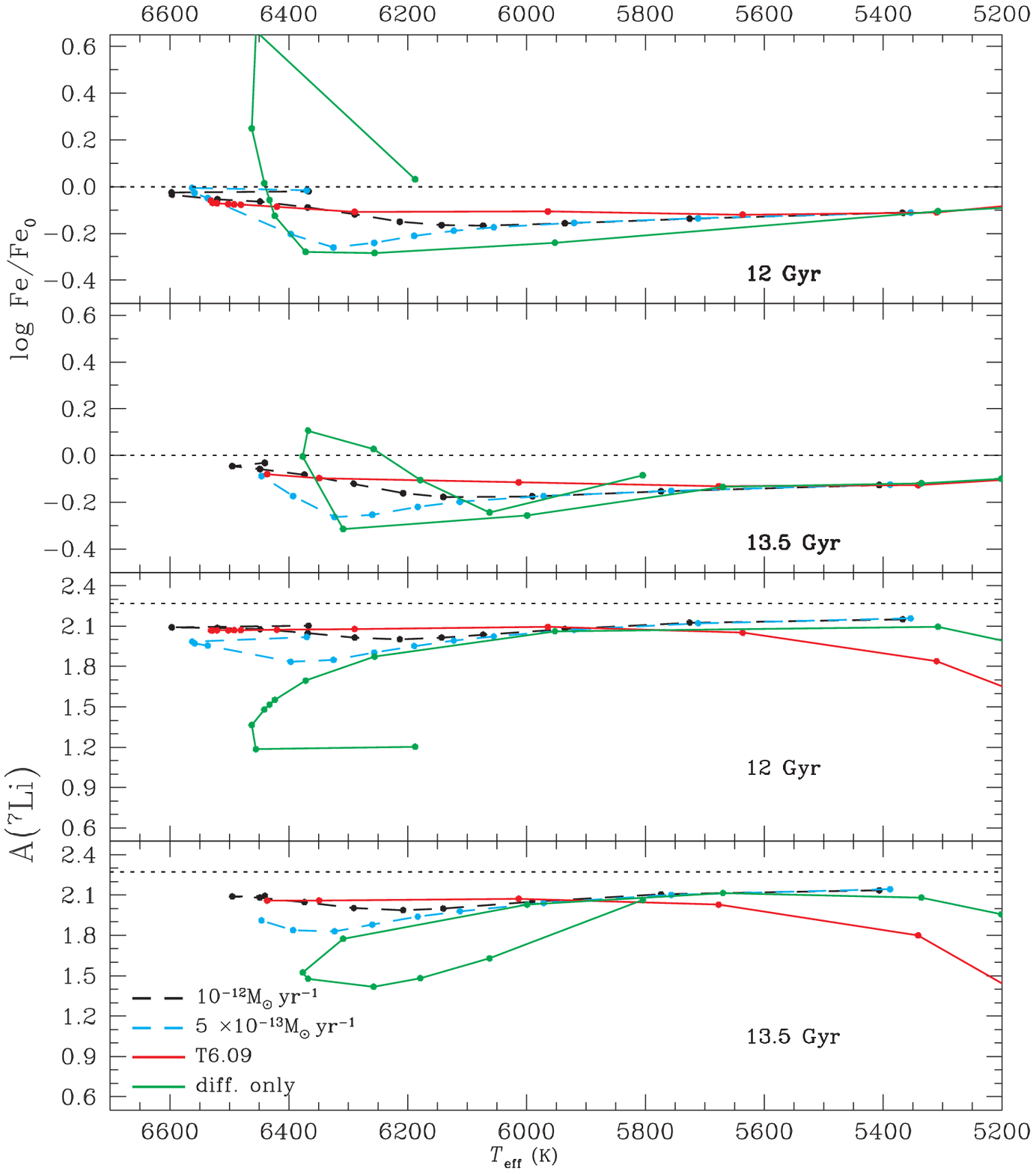}
\caption{Lithium and iron surface abundance isochrones at 12\,Gyr and 13.5\,Gyr for models with
and without mass loss as well as the T6.09 models with turbulence but no mass loss. The straight
line segments link  models (dots) calculated with the same physics. The horizontal dotted line at $A(^7\Li)=2.27$ 
indicates the initial lithium abundance for all models. The model masses 
are not the same for every isochrone. 
}\label{fig:spite}
\end{center}
\end{figure*}
Figure\,\ref{fig:spite} shows lithium and iron isochrones 
for the \textit{diffusion only} model, 
for models with mass loss rates of $10^{-12}\Mloss$ and $5\times 10^{-13}\Mloss$, as 
well as for T6.09
models\footnote{\label{fn:T609}The parameters specifying turbulent transport coefficients are
indicated in the name assigned to the model.  For instance, in the T6.09D400-3 model,
 the turbulent diffusion coefficient, $D\T$,  is 400 times greater
than the He atomic diffusion coefficient at $\log T_0 = 6.09$ and varying as $\rho^{-3}$, or
\begin{equation}
  \label{eq:DTT}
 \Dturb=400 D_{\mathrm{He}}(T_0)\left[\frac{\rho}{\rho(T_0)}\right]^{-3}. 
\end{equation}
 To simplify writing, T6.09 is also  used
instead of \mbox{T6.09D400-3} since all models discussed in this paper have the D400-3 parametrization. The 0.80T6.09 model is a 0.80\Msol{} model with T6.09 turbulence. } at 12\,Gyr and 13.5\,Gyr. To simplify the discussion only the effect of mass loss is considered in this section. The T6.09
isochrones that are  from \citet{richard02I}  are discussed in Sect.\,\ref{sec:turbulence} and where the effects of mass loss and turbulence are compared and the meaning of T6.09 is briefly described.

 In the \textit{diffusion only} model, large Fe and 
Li anomalies
are obtained around  turnoff: lithium decreases by a 
factor of up to 10 at 12\,Gyr, while Fe becomes 
overabundant by a factor of 5. For the mass
loss rate of $5\times 10^{-13}\Mloss$, the maximum lithium 
underabundance is reduced to -0.35\,dex, whereas
Fe overabundances are transformed into underabundances by factors 
below \hbox{-0.25\,dex}. For the 
$10^{-12}\Mloss$ model, the maximum amplitudes are further reduced. While 
the anomalies in the \textit{diffusion only} model near
turnoff are much smaller at 13.5\,Gyr than 
at 12\,Gyr, this effect is not
as important for lithium, and is unnoticeable for both models with 
mass loss. 

Below $\sim$6000\,K, the lithium isochrones with and without mass loss 
differ only slightly for preturnoff stars---with lithium depletions between 
-0.12 and -0.14\,dex near $\teff=5400\,$K---since the SCZ is 
too deep for mass loss to
have an effect. Though all three isochrones have decreasing lithium abundances  
for models hotter than $\sim$6000\,K, the underabundance amplitude decreases 
as the mass loss
rate increases. At 12\,Gyr, the lithium abundance begins increasing again 
near 6200\,K and 6400\,K for the models with $10^{-12}\Mloss$ 
and $5\times 10^{-13}\Mloss$
respectively. This is due to SCZ retraction, which exposes layers 
for which $v_{{\rm wind}}$ is dominant. 

The \textit{diffusion only} lithium isochrones do not exhibit 
plateau-like behavior. On the other hand, the 
mass loss isochrones are much flatter: the $5\times 10^{-13}\Mloss$ model 
has a scatter of about
$\sim$0.1\,dex and the $10^{-12}\Mloss$ model has a scatter 
of about $\sim$0.05\,dex with  
respective plateau average values of about --0.3\,dex and --0.2\,dex 
below the initial $^7$Li
abundance.

\section{Comparison to previous calculations with mass loss}
\label{sec:vauclair}
\begin{figure*}[!t]
\begin{center}
\includegraphics[scale=0.45]{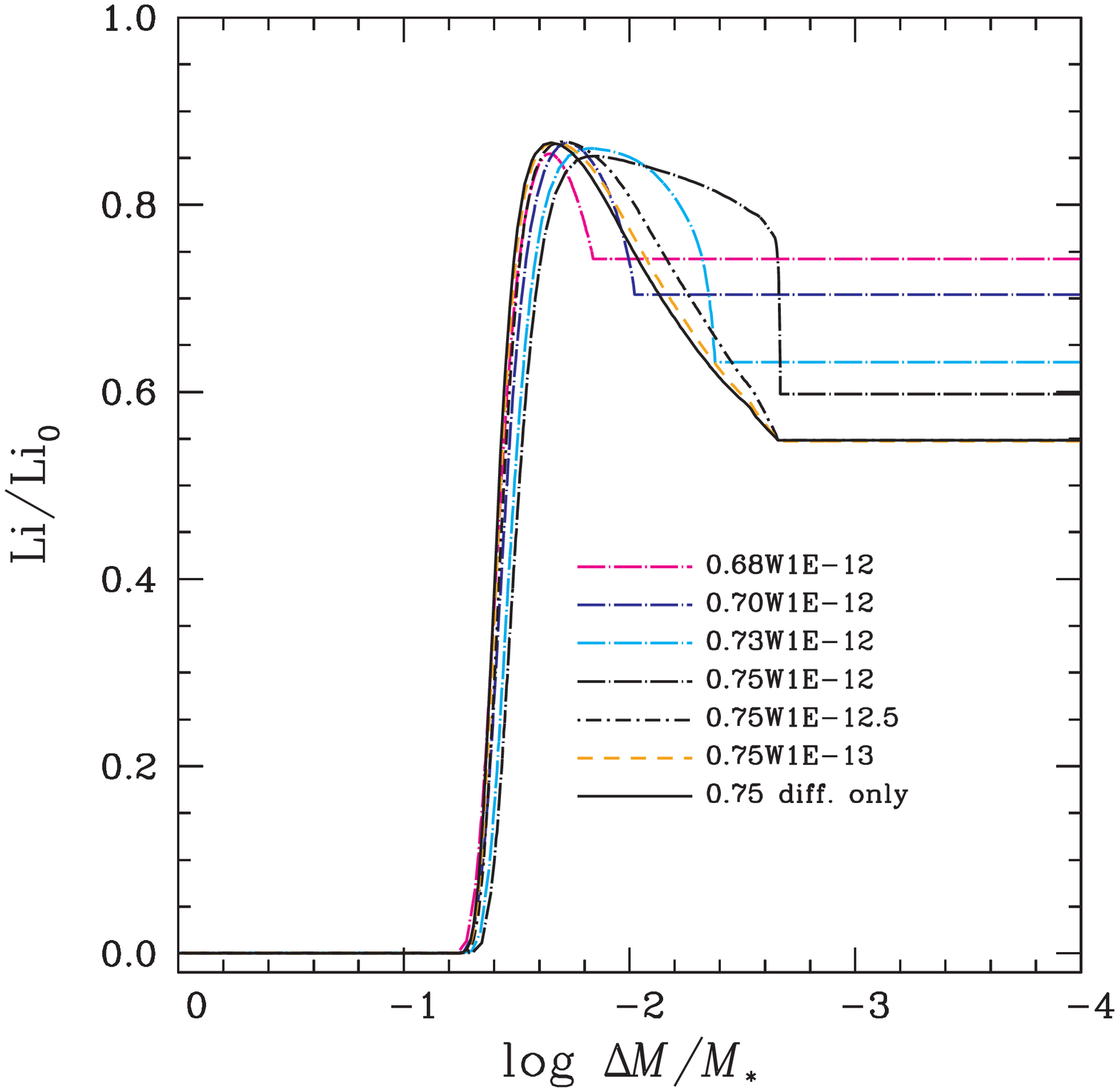}
\includegraphics[scale=0.45]{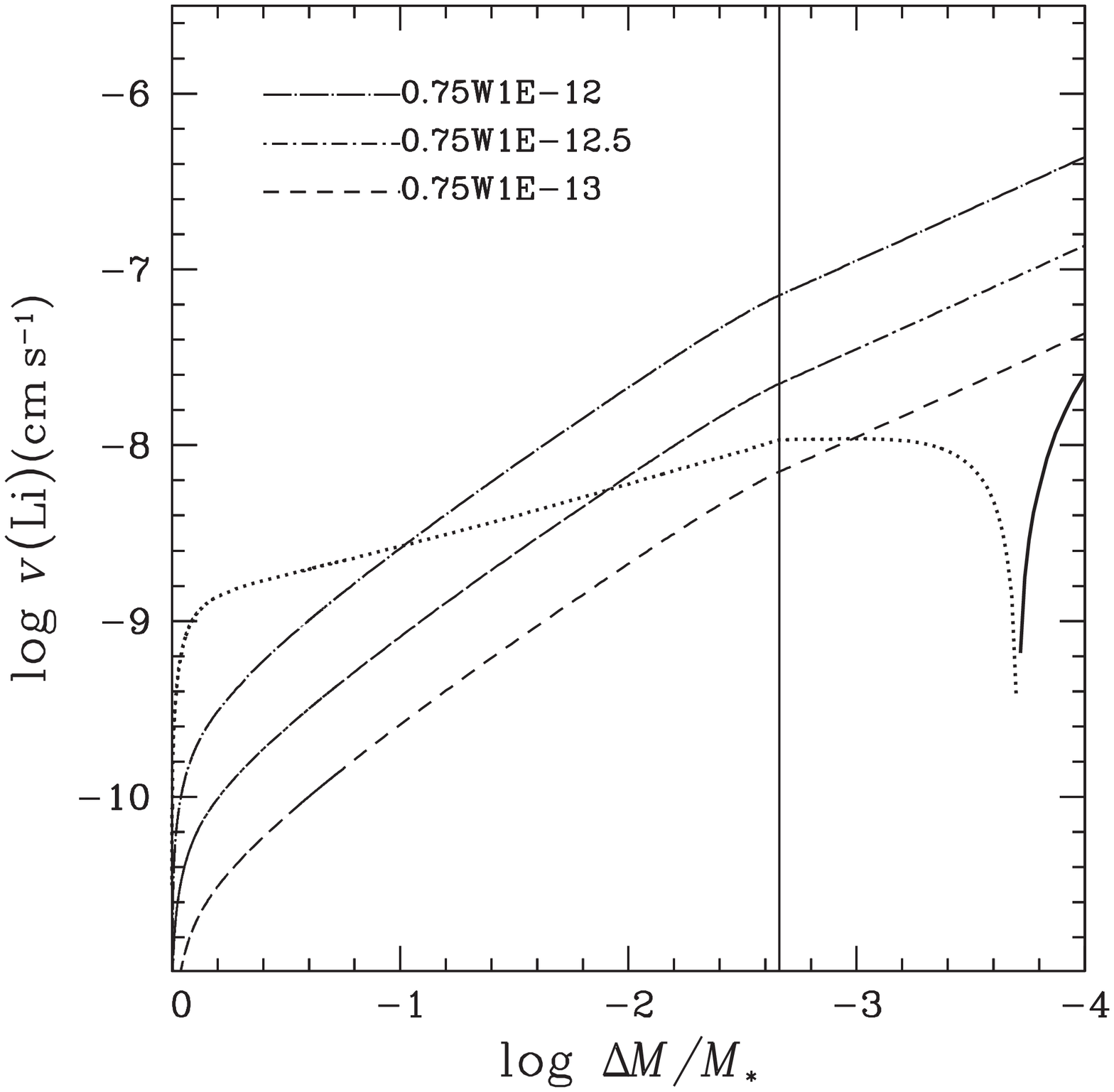}
\caption{[Left panel] Internal lithium profiles for 0.75\Msol{} models 
with and without mass loss. Three models of different masses were added (colored curves) 
in order to verify the solution for the 0.75W1E-12 model. [Right panel] A 
comparison of diffusion (dotted and solid lines) and wind velocities (dashed lines identified in the figure) for the 0.75\Msol{} models. The vertical solid line identifies the bottom of the surface convection zone. All models
are shown at 10\,Gyr. The left panel can be compared to Fig.\,4 of \citet{vauclair95}. }
\label{fig:vauclair}
\end{center}
\end{figure*}
In previous evolutionary calculations for Population II stars with mass loss, 
\citet{vauclair95} (VC95) found
that both mass loss rates of $10^{-12.5}$ ($\simeq 3\times 10^{-13}$) and 
$10^{-12}\Mloss$ lead to lithium abundances that 
increase monotonously (by at most 0.5 dex for the lower mass loss rate) 
with \teff{} for six models ranging from 0.60 to 0.80\Msol. They suggested
that the higher mass loss rate completely wipes out the effects of 
atomic diffusion, and leads to
a primordial lithium abundance of $A(\Li)=2.5\pm0.1$, which they determined from 
the surface abundance of their hottest model. 

Unfortunately, a detailed comparison to their results is impossible, 
since, even without mass loss, the stellar models are too different. These differences include $Y_0$, $Z_0$,  
$\alpha_{{\rm MLT}}$, the surface boundary condition, and 
the diffusion of metals. As a result, 
the SCZ evolution is very different. For instance, by comparing 
the internal lithium profile of the 
\textit{diffusion only} model in Fig.\,\ref{fig:vauclair} with the one from 
Fig.\,4 of \citet{vauclair95}, one sees that at 10\,Gyr, 
the SCZ (i.e. the flat part of the profile for both solutions) is about four times more massive 
in their models. Furthermore, their 0.75\Msol{} models are still on the MS 
(with their \teff{} still increasing) at 18\,Gyr, while our 0.75\Msol{} models 
have already
evolved toward the subgiant branch at 15\,Gyr. The difference 
in $Y_0$ is not large enough to
explain this significant discrepancy\footnote{\label{fn:Y2484}Though not shown, 
this was verified by evolving models with 
$Y_0=0.2484$, whose lower H mass fraction in the core led to their passing turnoff
only 1\,Gyr earlier than models with $Y_0=0.2352$. The models from 
\citet{vauclair95} have $Y_0=0.2403$}. 

That being said, some general 
results can still be compared. For instance, at 10\,Gyr, VC95 found that the 
surface lithium abundance for the 
$10^{-12.5}\Mloss$ model was 0.12\,dex larger than in the 
$10^{-12}\Mloss$ model, since the stronger mass loss rate reaches into 
the lithium burning layers. Our 
models also suggest that the strongest wind drags some material
from the lithium depleted layers
outward. Indeed, from Fig.\,\ref{fig:vauclair}, the wind velocity 
for the $10^{-12}\Mloss$ model at 10\,Gyr dominates the settling velocity down to
$\DM=-1$, the point of separation, which is well into the burning layers. However, 
at 10\,Gyr, this has not yet 
reached the surface since at $10^{10}$yr, a mass loss rate of $10^{-12}\Mloss$ has 
only exposed to the surface layers which are above $\sim 10^{-2}M_*$. In the same way, 
mass loss advected all mass above the point of separation, 
thus explaining the difference in the depths at which Li falls to zero in the 
\textit{diffusion only} and $10^{-12}\Mloss$ models. 

Furthermore, the shape of the internal
solutions in both sets of models is very different; for $10^{-12}\Mloss$, 
VC95 obtain a nearly flat solution from the BSCZ down to
Li--burning layers (see their Fig.\,4), whereas our solution suggests an important atomic 
diffusion--induced gradient below the SCZ (see Fig.\,\ref{fig:vauclair}). In order to verify 
that the difference was not simply
due to differences between  SCZ masses during evolution, this result was compared to smaller mass models
with the same mass loss rate (the 0.73W1E-12, 0.70W1E-12 and 0.68W1E-12 models). 
Even in models with more massive SCZs, the abundance gradient remains, since 
it is formed below $\DM\sim-2$, the point 
above which the wind has imposed the conservation of flux. 
The reason for these 
differences between the two calculations is not evident.

The \Li{} profiles in the left panel of Fig.\,\ref{fig:vauclair} are understood using the right panel and the assumption of unseparated mass loss from the surface.  In the 0.75\Msol{} diffusion only model, the gradient between $\DM = -1.8$ and --2.7 is determined mainly by the increasing settling time scale as \DM{} increases, that is closer to stellar center. In the 0.75W1E-13 model the wind only shifts $10^{-3}$ of a solar mass in $10^{10}$ years which only very slightly modifies the profile around $\DM \sim -2$.  In the 0.75W1E-12.5 model the effect is slightly more pronounced as the mass loss velocity becomes a little larger, in magnitude, than the diffusion velocity as seen in the right panel.  In the 0.75W1E-12 model however the mass loss velocity is larger than the diffusion velocity by a factor of up to six, and flux conservation determines the Li abundance between $\DM = -1.8$ and --2.7 as discussed in detail in \S\,5.1.1 of \citet{vick10}.  Since the mass loss from the surface  in unseparated  and the outgoing Li concentration is the \Li{} concentration in the convection zone, this implies no separation in the convection zone and the \Li{} mass loss flux is   $X(\Li) \rho  v_{\rm{wind}}$. It is as if the diffusion velocity were zero there.  Below the convection zone the \Li{} flux  is $\sim X(\Li) \rho(v_{\rm{wind}} -v_{\rm{D}})$.  Element conservation at the BSCZ forces a strong gradient just below\footnote{More technically, the diffusion term in $v_{\rm{D}}$ must compensate for the decrease in $X(\Li)$ as one approaches the convection zone.  Except where there is a strong $X(\Li)$ gradient, only the advective part of $v_{\rm{D}}$ plays a significant role.  Numerical accuracy required some one hundred zones in the small region of the strong gradient.}. It is seen mainly in the 0.75W1E-12 and 0.73W1E-12 models.  Because the wind velocity hardly exceeds the diffusion velocity below the bottom of the SCZ in the 0.68\,\Msol{} model, the gradient is less abrupt in that model.  

Finally, while VC95 obtain \textit{no} effects of chemical separation in 
all models with $10^{-12}\Mloss$ (see the caption of their Fig.\,4), lithium settling leads to 
$\sim$0.15\,dex surface
abundance reductions (at 12\,Gyr) in \textit{all} our models (see Fig.\,\ref{fig:spite}). 
In terms of isochrone shape, VC95 obtain a monotonous linear 
increase from the cooler stars 
to the hotter ones 
(Figs. 5 and 6 of VC95), while 
our isochrones dip in the interval 6400\,K$\,\geq \teff \geq\,$6000\,K (Fig.\,\ref{fig:spite}).

Some of the differences are related 
to the difference in models, namely the $\Delta M_{\rm BSCZ}$ 
evolution. 
Other differences could include the way the boundary condition below the SCZ was written and the fact that the wind term 
in their diffusion 
equation does not include
the often overlooked flux factor ($m_r/M_*$) which is required 
in evolutionary models (see Eq.\,[6] of Paper I). 
 This factor changes $v_{\rm{wind}}$ by 1\,\% at $\DM = -2$ and by 10\,\% at $\DM = -1$. Given that wind and diffusion velocities are comparable for many elements near $\DM = -1$ (see Figs. 3 and 4), this term modifies central concentrations.   For instance, in the 0.779\,\Msol{} model with a mass loss rate of $5 \times 10^{-13}\Mloss$, the central hydrogen concentration decreases below $10^{-3}$, $\sim 1.3$\,Gyr earlier when the correction term is included. The effect on surface concentrations is, however, very small.

\section{Comparison to models with turbulence}
\label{sec:turbulence}
Along with describing the effects of mass loss on Population II evolutionary models
with mass loss, 
this paper's main objective is to contrast models with mass loss to 
the ones with turbulence presented
in \citet{richard02I}. In particular, one goal is to see if a specific mass loss rate 
can reproduce the surface abundances exhibited in the T6.09 models$^{\ref{fn:T609}}$ around turnoff, which are compatible with
the Spite plateau (\citealt{RichardMiRi2005}).

In Fig.\,\ref{fig:absurfmdot}, the surface abundances of the 0.80T6.09 and 0.80W1E-12 models are shown to be 
very similar. Abundances of $^4\He$ and 
$^7\Li$ are about 0.02\,dex smaller in the model with turbulence, whereas elements with 
$Z>12$ are on average about 0.03-0.04\,dex 
smaller (with a maximum of about 0.05\,dex for Ar). The differences in surface abundances of 
elements from Be to Mg are smaller than the line width.

In terms of lithium and iron isochrones, as shown in Fig.\,\ref{fig:spite}, 
the differences are just as small. Both types of models lead to nearly flat abundances 
(above $\teff=5800$K), and their plateau value is nearly the same. At 12\,Gyr, the maximum 
difference in Fe abundance, which occurs near $\teff\simeq 6100$K, is about 0.07\,dex, though
this could be due to the absence of turbulent models in this region. Similarly, 
for models hotter than $\teff=5800$K, the maximum difference in $^7$Li 
abundances, which occurs at $\teff\simeq 6200$K, is  0.1\,dex, and could again be explained by
the T6.09 isochrone's poor resolution. Below $\teff=5800$K, larger differences appear for lithium, likely 
due to lithium destruction in the model with turbulence. The discrepancy is greater than 0.3\,dex, although again, the resolution is not
sufficient to quantify precisely the difference.   

\begin{figure*}[!t]
\begin{center}
\includegraphics[scale=1.0]{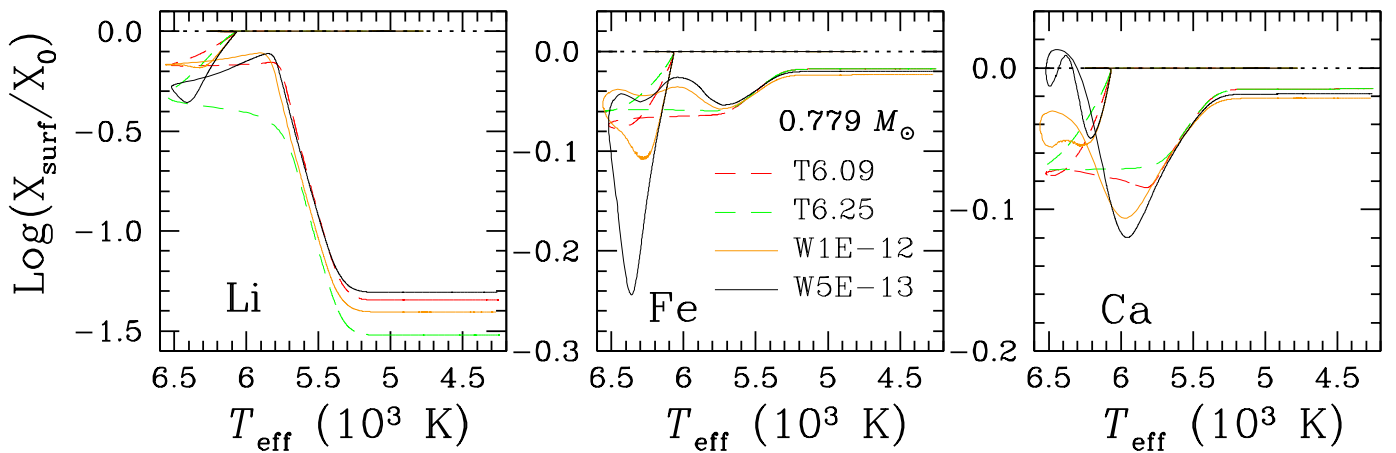}
\caption{Evolution of the surface  abundances of Li, Fe, and Ca as a function of \teff{} for two models with turbulence and two with mass loss  identified in the figure. The chosen scale emphasizes the behavior on the subgiant and bottom of the giant branch. All models start (pre--main--sequence) to the right of the upper horizontal line and evolve to the left until they reach the main sequence at $\teff \sim 6100$\,K.  Then, diffusion starts being effective and the Li abundance, for instance, is reduced by 0.2 to 0.4 dex as the star approaches turnoff, at $\teff \sim 6600$\,K.  The models then move to the subgiant and giant branches where dilution further reduces the surface Li abundance by a factor of $\sim 10$.  For \Ca{} and \Fe{} the radiative acceleration complicates the surface abundance evolution in the  mass loss cases.}
\label{fig:SGB} 
\end{center}
\end{figure*}
Whereas the surface behavior is quite similar for most 
models with T6.09 and $10^{-12}\Mloss$, 
the internal solutions are a little more different. By comparing Fig.\,\ref{fig:temporel} with 
Fig.\,8 of \citet{richard02I}, one notes that in the 0.80W1E-12 model,
abundances change gradually from the bottom of the SCZ down to about $\DM=-1$, with most 
variations between
0.1 and 0.15\,dex. However, in the 0.80T6.09 model at turnoff, abundances are 
constant from the surface down to just below $\DM=-2$. 
Even though in the mass loss
models the abundance variations
are relatively small, 
and  the corresponding changes in opacity are likely to be 
small as well, this could potentially
be tested by asteroseismic observations.

\section{Subgiant and giant branches}
\label{sec:Subgiant}
The evolution of the surface abundances of \Li{}, \Fe{}, and \Ca{} are shown in Fig.\,\ref{fig:SGB} for  0.779\,\Msol{} models with $Z_0 = 0.00027$\footnote{As is appropriate for NGC\,6397.  As stated in \S\,\ref{sec:calcul} and \ref{sec:models}, most other models have $Z_0 = 0.00017$.}.  It is shown as a function of \teff{} in order to facilitate comparison with stars on  the subgiant and giant branches. 
At turnoff, the T6.25 model has approximately the same Li abundance as the $5 \times 10^{-13}\Mloss$ model and the T6.09 model approximately the same as the $10^{-12}\Mloss$ model.  Past turnoff, the surface \Li{} abundance first increases in both models with mass loss (with a maximum at $\teff \sim 5800$\,K) as the peak in Li abundance seen at $\DM \sim -2$ in the left panel of Fig.\,\ref{fig:vauclair} is mixed to the surface.  In the models with  turbulence, this Li peak is either reduced (T6.09 model) or eliminated (T6.25 model) as may be seen in Fig.\,7 of \citet{RichardMiRi2005}. In the latter model, the Li abundance decreases continuously past turnoff to {$\sim -1.5$ on the giant branch ($\teff < 5000$\,K). More Li has been destroyed by turbulence in the T6.25 model than carried away by a mass loss rate of $10^{-12}\Mloss$ since the Li abundance on the giant branch is smaller in the T6.25 model than in the mass loss one.  This implies that turbulence leads to a 0.2\,dex additional destruction of Li in the T6.25 model when compared to models with atomic diffusion alone.    

These results may be compared to the calculations and observations of \citet{MucciarelliSaLoetal2011} and of \citet{NordlanderKoRietal2012}. The T6.25 model leads to an additional 0.2\,dex destruction of Li  as appears required for the observations of Li in lower red giant branch stars \citep{MucciarelliSaLoetal2011} to be compatible with WMAP measurements.  Neither of the mass loss models leads to as large a reduction of surface Li on the giant branch as the T6.25 model, so that the RGB observations favor the turbulent T6.25 model.  However, one may also compare our results to the Li observations in Figs.\,2 and 5 of \citet{NordlanderKoRietal2012} where isochrones of Li abundance are compared to observations of NGC\,6397. While we do not have  isochrones, a comparison is possible using evolutionary variations of Li abundance since the same turbulent models were used in our calculations and for their Figs.\,2 and 5. If one compares the $10^{-12}\Mloss$ results, to the T6.0 preferred by \citet{NordlanderKoRietal2012}, one notes similar agreement at turnoff, but better agreement on the subgiant branch since maxima in our Fig.\,\ref{fig:SGB} appear at a slightly higher \teff{} and are more pronounced for the $10^{-12}\Mloss$ model than for the two turbulent cases.  The agreement is also better for the bRGB and RGB observations since the $10^{-12}\Mloss$ curve is lower than even the T6.09 curve in our Fig.\,\ref{fig:SGB}.  It then appears that models with mass loss lead to at least as good agreement with Li observations as do models with turbulence.

For \Ca{} and \Fe{}, the radiative accelerations complicate the evolution in the mass loss cases.  For instance, the large \Ca{} abundance close to the surface for the 11.664 Gyr curve for \Ca{} on Fig.\,\ref{fig:temporel} is caused by the \gr(\Ca) minimum as discussed in relation to that figure.  In turn it appears as the abundance maximum at $\teff \sim 6500$\,K  in Fig.\,\ref{fig:SGB}.  The \Fe{} local maximum at $\teff \sim 6000$\,K in  Fig.\,\ref{fig:SGB} is similarly caused by the \gr(\Fe) minimum for the 11.664 Gyr curve at $\DM \sim -3$ in Fig.\,\ref{fig:temporel}.  These properties affect the comparison to observations for the mass loss models.  The comparison is shown for the turbulent models in Fig.\,2 of \citet{NordlanderKoRietal2012}.  If one assumes the abundance shifts just discussed, the agreement with observations would be  poorer with the mass loss models for both \Fe{} and \Ca{} at $\teff \sim 6400$\,K while at $\teff \sim 6000$\,K, it would be  poorer for \Fe{} and unchanged for \Ca{}.

 Given the difficulty  justifying such a high mass loss rate we did not push the comparison with observations  further.

}
\section{Conclusion}
\label{sec:conclusion}

Evolutionary models with atomic diffusion and unseparated mass loss explain some
characteristics observed in galactic halo dwarfs. A mass loss rate 
of $5 \times 10^{-13}\Mloss$ leads to surface lithium isochrones that 
exhibit plateau-like behavior with a maximum scatter of about 
0.1\,dex around an average
value that is -0.3\,dex below the initial lithium abundance (Fig.\,\ref{fig:spite}). 
For models with mass loss rates of $10^{-12}\Mloss$, 
the lithium isochrones are even flatter with a scatter of about 0.05\,dex around
an average value about -0.2\,dex lower than the initial abundance.
 
While mass loss does not have a significant impact on a star's structural properties, its
effects on abundances, both in the interior and on the surface, can be important. 
In the interior, abundance profiles are either dominated by atomic diffusion---and reflect
local variations in \gr---or are dominated by the wind---and adjust 
in order to conserve the
flux coming from below (Sect.\,\ref{sec:abint}). A mass loss rate 
of $10^{-13}\Mloss$ reduces both 
internal and surface anomalies by up to a factor of 6-7, while a mass loss rate 
of $10^{-12}\Mloss$ reduces all anomalies to below 0.2\,dex (Fig.\,\ref{fig:mdot} and \ref{fig:absurfmdot}). 
While the $10^{-12}\Mloss$ models lead to
generalized underabundances for all elements, lower mass loss rates allow for
overabundances to develop at the surface.

The mass loss models presented in this paper differ significantly from those 
presented in \citet{vauclair95} (Fig.\,\ref{fig:vauclair}). 
Although some 
of this difference can be attributed to
differences in input parameters, other discrepancies 
are more difficult to explain (Sect.\,\ref{sec:vauclair}). In one critical aspect however, the results are robust: both calculations conclude that below $10^{-14}\Mloss$ mass loss has virtually no effect while above $10^{-12}\Mloss$ it eliminates all effects of atomic diffusion from the surface.

With respect to the T6.09 models from \citet{richard02I}, the models with a mass loss rate of $10^{-12}\Mloss$
are very similar above $\teff=5800$K (Figs.\,\ref{fig:absurfmdot} and \,\ref{fig:spite}). 
Below this temperature, turbulence leads to greater lithium destruction. 
Because of differences in the internal distribution of elements 
(Sect.\,\ref{sec:turbulence}), asteroseismology could perhaps distinguish between the 
two scenarios.
 
Unfortunately, even for mass loss rates of $10^{-12}\Mloss$, the implied stellar winds  are barely detectable through direct observation. 
Furthermore, considering that acceleration processes for these types of winds are
unknown, it is difficult to justify mass loss rates that are 50 times higher than the
solar mass loss rate ($\sim 2 \times 10^{-14}\Mloss$). It is unlikely that winds in Population II
stars are driven by radiation; therefore, if these stars host solar-like winds, 
one could hope to observe solar-like coronae and/or X-ray emission.

A signature of a chromosphere has recently been observed in Pop II dwarfs: the \He{} line 10\,830\,A is present in almost all dwarfs with [Fe/H] $= -0.5$ to $-3.7$ \citep{TakedaTa2011}.  Signs of chromospheres had earlier  been reported in these stars by \citet{PetersonSc2001}.  On the other hand, \citet{CramerSa2011} have recently evaluated  mass loss rates in various types of stars.  According to their Fig.\,13, mass loss rates of  $\sim 10^{-12}\Mloss$ are to be expected in dwarf stars of $\teff \sim 6000$\,K of solar metallicity that have rotation periods of around six days.  They do not give equivalent values for lower metallicity stars. The most critical factor is, however, the rotation rate.  Pop\,II dwarfs are believed to rotate slowly.  However, there are observations (reviewed in  Fig.\,6 of \citealt{TakedaTa2011}) suggesting that many rotate at $v_e \sin i \sim 5$\,km/s.  It would seem important to confirm both the rotation rate observations and expected mass loss rates for low metallicity stars.  In the meantime it seems premature to disregard mass loss as a process competing with atomic diffusion in Pop\,II stars.

As a next step, these models could be extended to the He flash in order to further constrain the effects of
mass loss in Population II giants, for which mass loss has extensively been observed and
studied (\citealt{origlia07}). It would also be interesting to see if such
mass loss rates would affect, or even eliminate, the effects of atomic diffusion 
(\citealt{michaud10}) and/or of
thermohaline mixing (\citealt{charbonnel07,TraxlerGaSt2011}).


\acknowledgements{M.Vick would like to thank the D\'epartement de physique de 
l'Universit\'e de Montr\'eal for financial support, along with
 everyone at the GRAAL in Montpellier for their amazing hospitality. 
We acknowledge 
the financial support of Programme National de Physique 
Stellaire (PNPS) of CNRS/INSU, France. This research was
partially supported by NSERC at the Universit\'e de 
Montr\'eal. Finally, we  
thank Calcul Qu\'ebec for providing us with the
computational resources required for this work. We thank an anonymous referee for useful comments.}



\Online
\begin{appendix} 
\section{Version of Figures\,\ref{fig:temporel} and \ref{fig:mdot} with all  atomic species included in the calculations.}
\label{sec:Appendix}

\begin{figure*}[!t]
\begin{center}
\includegraphics[scale=0.80]{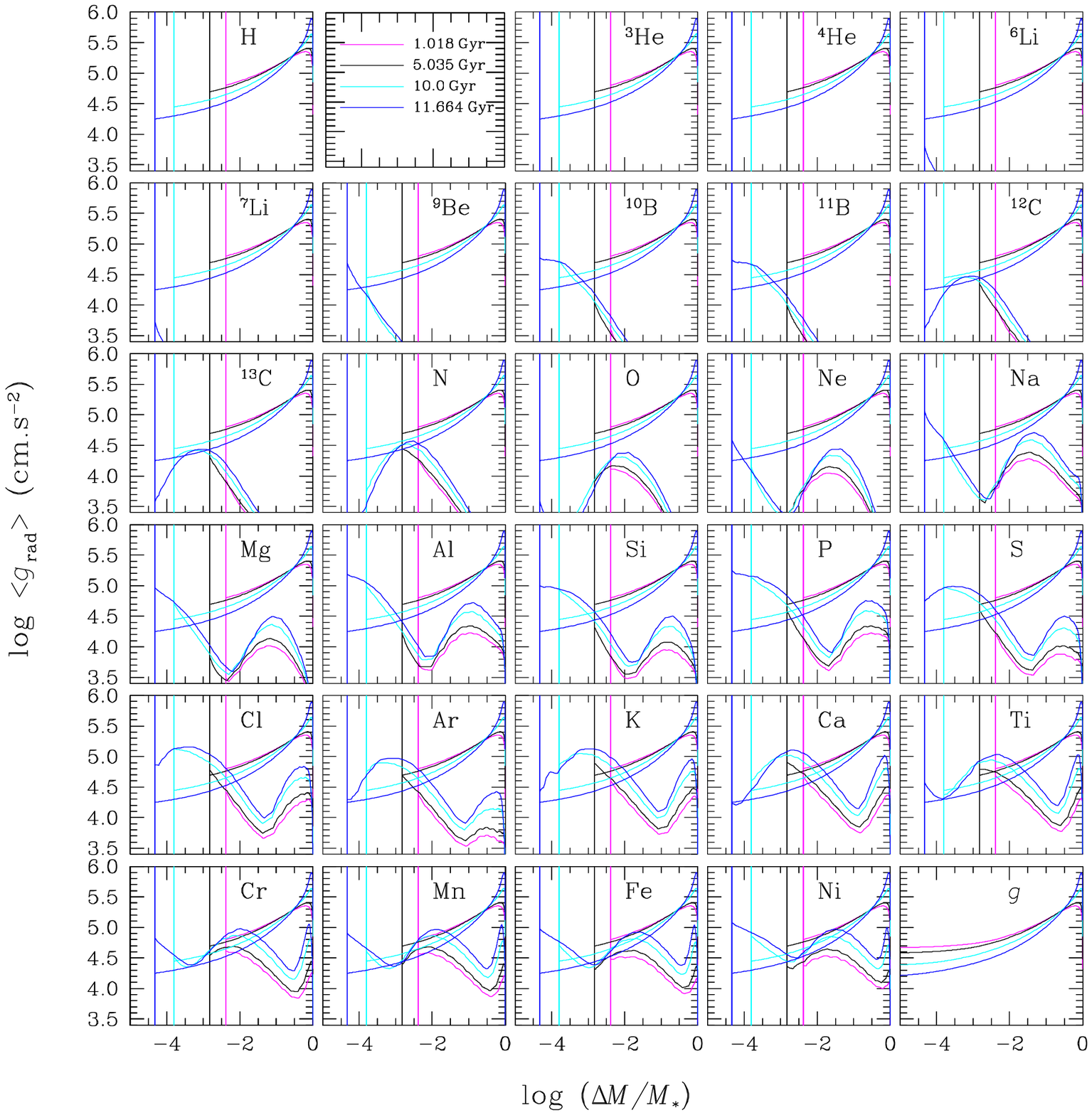}
\caption{ Radiative accelerations  and gravity  
at four 
different ages in a 0.8\Msol{} model with 
${\dot M}=10^{-12}\Mloss$. The vertical lines indicate 
the position of the bottom of the SCZ. The ages are identified in the figure.}
\label{fig:allGr}
\end{center}
\end{figure*}

\begin{figure*}[!t]
\begin{center}
\includegraphics[scale=0.80]{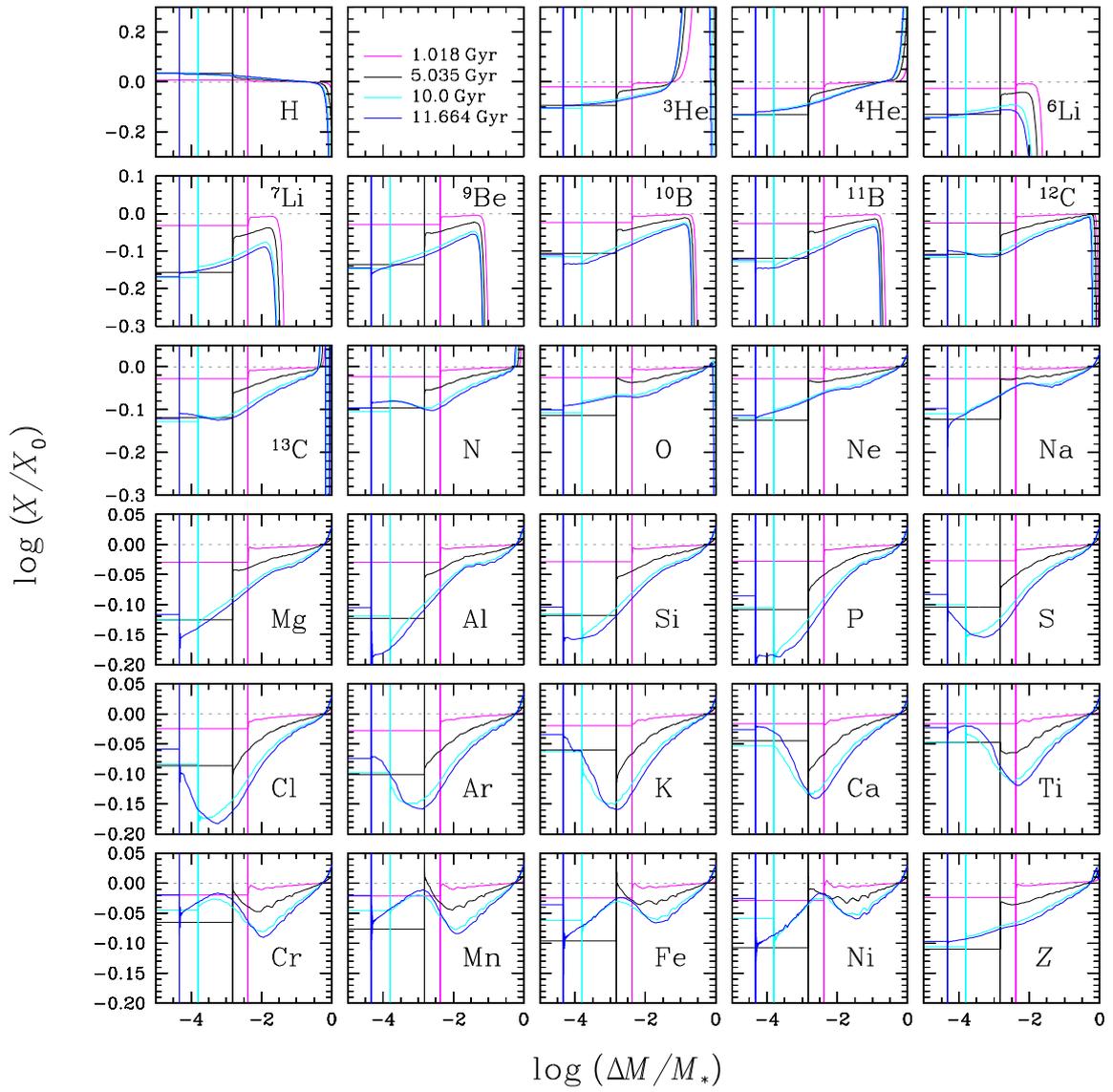}
\caption{ Abundance profiles
at four 
different ages in a 0.8\Msol{} model with 
${\dot M}=10^{-12}\Mloss$and an initial metallicty of $Z_0=0.00017$. The vertical lines indicate 
the position of the bottom of the SCZ. The ages are identified in the figure.}
\label{fig:allX_age}
\end{center}
\end{figure*}

\begin{figure*}[!t]
\begin{center}
\includegraphics[scale=0.80]{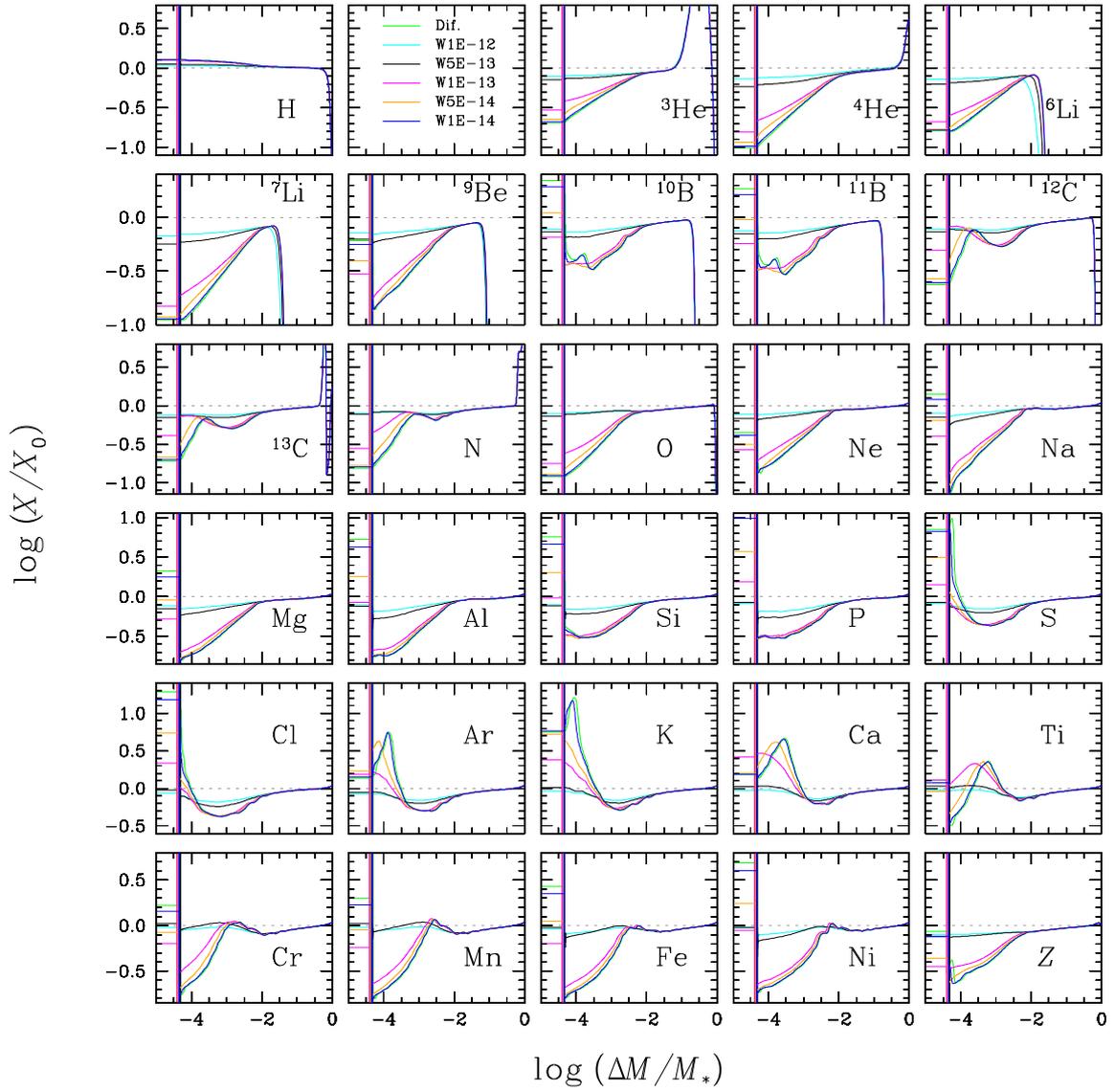}
\caption{ Abundance profiles for the model without mass loss and 
for  mass loss rates, from ${\dot M}=10^{-14}\Mloss$ to  ${\dot M}=10^{-12}\Mloss$ in a 0.8\Msol{} model and an initial metallicty of $Z_0=0.00017$.  The internal concentration profiles are shown just before 
turnoff near 11.5\,Gyr.  The vertical lines indicate 
the position of the bottom of the SCZ. The mass loss rates  are identified in the figure.}
\label{fig:allX_Mloss}
\end{center}
\end{figure*}

\begin{figure*}[!t]
\begin{center}
\includegraphics[scale=1.0]{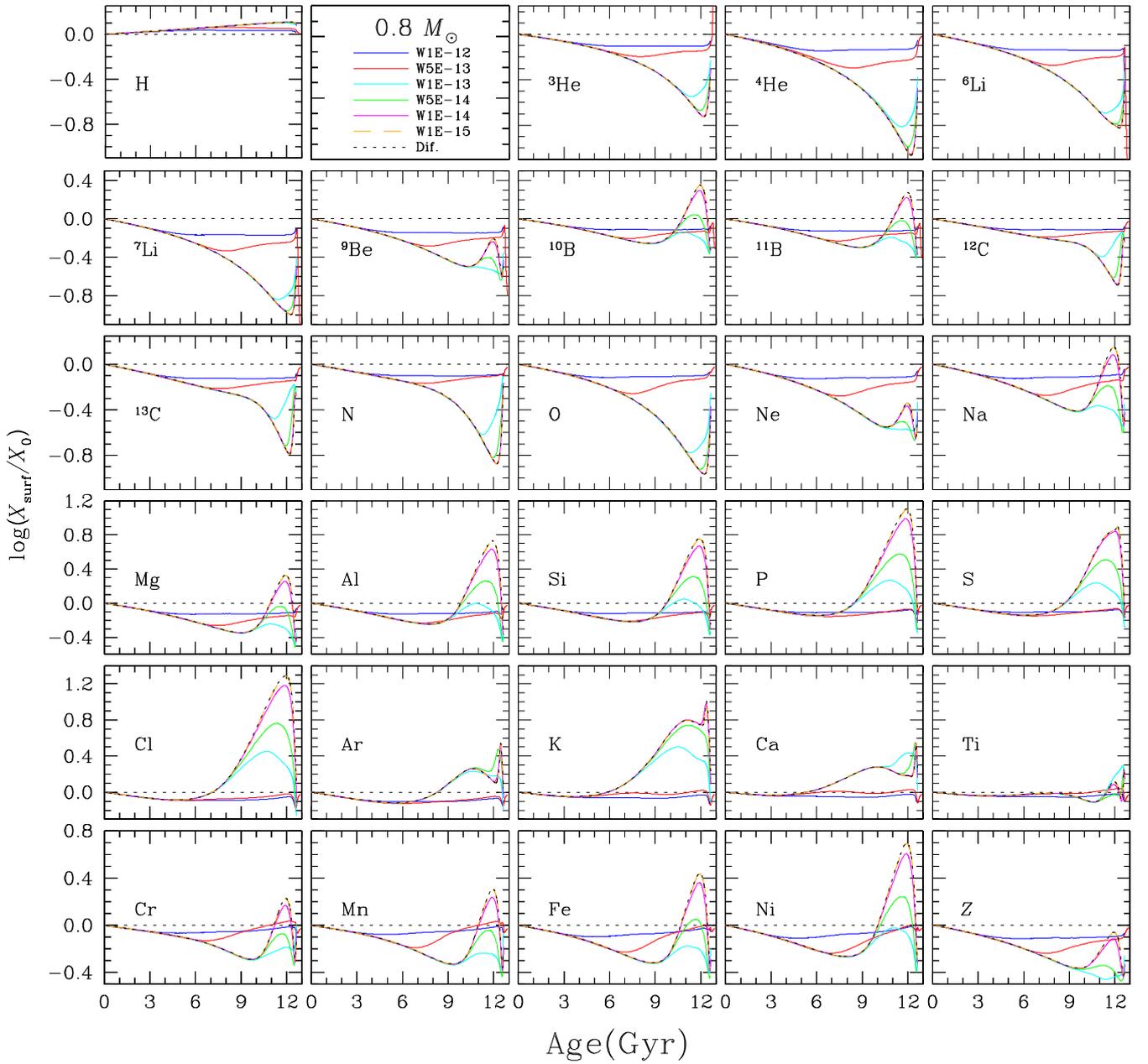}
\caption{ Evolution of surface abundances  for the model without mass loss (dotted line) and 
for  mass loss rates, from ${\dot M}=10^{-15}\Mloss$, to  ${\dot M}=10^{-12}\Mloss$ in a 0.8\Msol{} model and an initial metallicty of $Z_0=0.00017$. One cannot distinguish the curves for the model without mass loss from those for the model with  ${\dot M}=10^{-15}\Mloss$.  The mass loss rates  are identified in the figure.}
\label{fig:allX_evol_Mloss}
\end{center}
\end{figure*}

\end{appendix}
\clearpage
\end{document}